\documentclass[twocolumn, nofootinbib, aps, prl, floats, floatfix, amsmath, amssymb, longbibliography, superscriptaddress, preprintnumbers]{revtex4-1} %
\usepackage[final]{graphicx}
\usepackage{hyperref}
\usepackage{amsmath}
\usepackage{bbm}
\usepackage{bm}
\usepackage{amsfonts}
\usepackage{amssymb}
\usepackage{latexsym}
\usepackage{graphicx}
\usepackage[english]{babel}
\usepackage{multirow}
\usepackage{float}
\usepackage{url}
\usepackage{slashed}
\usepackage{xcolor} 
\usepackage[utf8]{inputenc}
\usepackage{stmaryrd} 
\usepackage{enumitem}
\usepackage{hyperref}
\usepackage{cleveref}
\usepackage{siunitx}
\usepackage{verbatim}
\usepackage{pifont}
\usepackage{tikz}

\usepackage{amsthm}

\theoremstyle{remark}

\newcommand{\eg}{e.g.,\ }	

\newcommand{\MeV}{{\rm MeV}} 
\newcommand{\GeV}{{\rm GeV}} 

\definecolor{darkerblue}{rgb}{0.2,0.2,0.5}
\definecolor{seagreen}{rgb}{0.180392,0.545098,0.341176}
\definecolor{smagenta}{rgb}{0.5,0.145098,0.341176}
\definecolor{deepblue}{rgb}{0,0,1}

\begin{document}
\title{New mechanism for primordial black hole formation from the QCD axion}

\author{Shuailiang Ge}
\email{sge@pku.edu.cn}
\affiliation{Center for High Energy Physics, Peking University, Beijing 100871, China}
\affiliation{School of Physics and State Key Laboratory of Nuclear Physics and Technology, Peking University, Beijing 100871, China}

\author{Jinhui Guo}
\email{guojh23@pku.edu.cn}
\affiliation{School of Physics and State Key Laboratory of Nuclear Physics and Technology, Peking University, Beijing 100871, China}
 
\author{Jia Liu}
\email{jialiu@pku.edu.cn}
\affiliation{School of Physics and State Key Laboratory of Nuclear Physics and Technology, Peking University, Beijing 100871, China}
\affiliation{Center for High Energy Physics, Peking University, Beijing 100871, China}

\begin{abstract}
We present a new mechanism for the primordial black hole (PBH) production within the QCD axion framework. We take the case where the Peccei-Quinn symmetry breaks during inflation, resulting in a $N_{\rm DW}=1$ string-wall network that re-enters horizon sufficiently late. Therefore, closed axion domain walls naturally arising in the network are sufficiently large to collapse into PBHs. Our numerical simulation shows that 
$\sim 0.3\%$ 
of the total wall area is in the form of closed walls. 
In addition, the relic abundance of dark matter is dominantly accounted for by free axions from the collapse of open walls bounded by strings. 
In this framework, the abundance of PBH within dark matter is calculated to be
$\sim 0.9\%$.
This fraction remains unaffected by axion parameters or the re-entering horizon temperature, as it is determined by the fixed proportion of closed walls in the network, governed by the principles of percolation theory.
The resultant PBHs uniformly share the same mass, which spans from about $10^{-9}$ to $1$ solar mass, corresponding to the classical QCD axion mass window $10^{-5}-10^{-2}$~eV and the re-entering horizon temperature $300-1$~MeV.
Intriguingly, PBHs in this mechanism can naturally account for the ultrashort-timescale gravitational microlensing events observed by the OGLE collaboration.
\end{abstract}
\maketitle

\noindent \textit{\textbf{Introduction}}--
QCD axion is an elegant solution to the strong CP problem via the Peccei-Quinn (PQ) mechanism~\cite{peccei1977cp, peccei1977constraints, weinberg1978new, wilczek1978problem, kim1979weak, shifman1980can, dine1981simple, Zhitnitsky:1980tq}, which is one of the most promising directions for physics beyond the Standard Model.

This mechanism introduces a complex scalar field $\Phi$ with a global $U(1)$ symmetry in a potential $V= \lambda/4 \cdot (\left| \Phi \right|^2 - v_a^2)^2$.  The symmetry spontaneously breaks when the cosmological temperature falls below the PQ energy scale $v_a$. Axion cosmology is usually studied in two scenarios, 
depending on the relation between $v_a$ and the inflationary Hubble scale $H_I$; see \eg Refs.~\cite{Sikivie:2006ni, Marsh:2015xka, DiLuzio:2020wdo} for a review. If $v_a \lesssim H_{I}/2\pi$, PQ symmetry breaks after inflation. 
Axion strings first form as the Goldstone field $\theta \equiv a/v_a$ of $\Phi$ winds around from 0 to $2\pi$ at boundary, where the angular variable $\theta$ is the rescaled dimensionless version of the axion field $a$. Later, when temperature drops to $T_{1} \sim \mathcal{O}({\rm GeV})$, axion mass becomes comparable to Hubble rate, $m_a(T_1)\sim H(T_1)$, and the axion effective potential 
from non-perturbative QCD explicitly breaks PQ symmetry down to a discrete $Z(N_{\rm DW})$ symmetry with $f_a = v_a/N_{\rm DW}$. Here, $N_{\rm DW}$ is the domain wall number. Consequently, axion domain walls form with the pre-existing strings as boundaries. 
However, for $v_a \gtrsim H_{I}/2\pi$, PQ symmetry breaks before inflation, and the formed strings are blown away out of the current Hubble horizon $H_0^{-1}$ by inflation. Thus, the axion field gets homogenized on the same scale, and the axion topological defects have no observable effects. 

In addition to the above post- and pre-inflationary scenarios, PQ symmetry can also break \textit{during} inflation. This can happen naturally if the PQ breaking is driven by inflation and can be achieved in various ways (\eg Refs.~\cite{Harigaya:2022pjd, Redi:2022llj}; see also Ref.~\cite{An:2023idh}).
For example, one can couple the PQ field $\Phi$ to the inflaton field $\phi$ via $c \phi^2 \Phi^{\dagger}\Phi$ where $c$ is the coupling. Then, during inflation, the excursion of $\phi$ from large field induces PQ breaking when $\phi$ rolls down to $\phi = \sqrt{\lambda/2c} v_a$. This example works for $v_a > H_I/2\pi$ where the finite temperature correction due to $H_I$ will not affect the breaking scale. Also, for $v_a$ much larger than $H_I$, the fluctuation $\delta \theta \propto H_I/v_a$ during inflation is greatly suppressed~\cite{Redi:2022llj}, which will not disturb the correlation length of topological defects.

In the during-inflationary scenario, the string-wall network re-enters horizon at temperature $T_{\rm en}$. Supposing that PQ breaks after $N_{\rm PQ}$ e-foldings of visible inflation, we have
$T_{\rm en} \simeq T_0 {\rm e}^{N_{\rm PQ}+4}$~\cite{Redi:2022llj} 
where $T_0$ is the current cosmological temperature.  
$T_{\rm en}$ should be higher than $\sim 1$~MeV to avoid possible violations of BBN processes. On the other hand, $T_{\rm en}$ should be smaller than $T_{1}\sim 1~\GeV$, such that walls form when strings are still super-horizon. Otherwise, it makes no difference with the post-inflationary scenario. Since the network is super-horizon at $T_1$, it will not collapse until $T_{\rm en}$. 
Dominated by wall energy, the network collapses right after $T_{\rm en}$ as strings eat walls rapidly under the wall tension~\cite{Harigaya:2022pjd}. This holds for the case $N_{\rm DW} = 1$ which naturally avoids the domain wall problem. Since such collapse can be much later than that in the post-inflation scenario, it opens new parameter space for the resultant free axions accounting for dark matter~\cite{Harigaya:2022pjd, Redi:2022llj}. 

In addition to walls bounded by strings,  a small portion of the network is in the form of closed walls~\cite{Vachaspati:1984dz, Chang:1998tb}. Closed walls naturally arise due to the same Kibble-Zurek mechanism~\cite{Kibble:1976sj, Zurek:1985qw} when one vacuum in space is fully surrounded by another topologically different vacuum. Compared with the walls bounded by strings,  closed walls received less attention in the literature. However, closed walls can play an important role in the early universe by collapsing into primordial black holes (PBHs)~\cite{Vachaspati:2017hjw, Ge:2019ihf}.

In this study, we will show that in the during-inflationary scenario, the closed walls collapse into PBHs naturally for most of the parameter space after re-entering horizon at $T_{\rm en}$. Closed walls are big enough because $T_{\rm en}$ is sufficiently late. It overcomes the difficulty in Refs.~\cite{Vachaspati:2017hjw, Ge:2019ihf} of finding large closed walls for forming PBHs. Moreover, this mechanism differs fundamentally from the following ones, including Ref.~\cite{Ferrer:2018uiu} which requires the presence of false vacuum induced by a bias potential to form PBHs and requires $N_{\rm DW}\geq 2$, Ref.~\cite{Rubin:2000dq} which needs closed walls generated by fluctuations of axion field, and Refs.~\cite{Kitajima:2020kig, Kasai:2023ofh, Li:2023det, Li:2023zyc} which need the generation of overdensity fluctuations of particle axions to form PBHs.

In our scenario, PBH abundance in dark matter is $0.92\%\pm0.05\%$ and the remaining dark matter is accounted for by free axions from walls bounded by strings. This proportion is fixed, independent of axion parameters or $T_{\rm en}$, because the proportion of closed walls in the network is fixed, which can be determined by numerical simulations. 
The resultant PBH mass is monochromatic, which can be $10^{-9}-1 ~ M_{\odot}$ for the classical QCD axion mass window $10^{-5} - 10^{-2}$~eV.
We want to emphasize that apart from assuming that PQ symmetry breaks during inflation, we do not make any changes or fine-tunings to the standard QCD axion cosmology, which is remarkable that PBHs can form in this simple picture. Furthermore, such PBHs can potentially explain the gravitational microlensing events observed by the OGLE collaboration~\cite{Niikura:2019kqi, 2017Natur.548..183M} and also serve as a candidate for the intriguing Planet 9 in the solar system if it is interpreted as a PBH~\cite{Scholtz:2019csj, Witten:2020ifl}. 
\\

\noindent \textit{\textbf{Closed axion domain walls}}--
We are going to simulate the string-wall network formation for the case $N_{\rm DW}=1$, particularly focusing on the proportion of closed walls. First of all, we stress that a notoriously difficult simulation problem in the post-inflationary scenario does not arise here. The problem is essentially a multi-scale problem that strings and walls are formed at drastically different times, $f_a\gg T_1$, making it impossible for computers to \textit{fully} simulate the long-time string evolution before $T_1$ unless taking some compromises~\cite{Hiramatsu:2010yn, Hiramatsu:2012gg, Kawasaki:2014sqa, Hiramatsu:2012sc, Fleury:2015aca, Klaer:2017ond, Klaer:2017qhr, Vaquero:2018tib, Gorghetto:2018myk, Buschmann:2019icd, Gorghetto:2020qws, Hindmarsh:2021zkt}. In our scenario, strings are frozen in the super-horizon scale without dynamics. Therefore, it is equivalent to treating walls and strings as forming simultaneously in the sense that string evolution is trivial before wall formation.  

\begin{figure}[tb]
    \centering
     \includegraphics[width=0.8\linewidth]{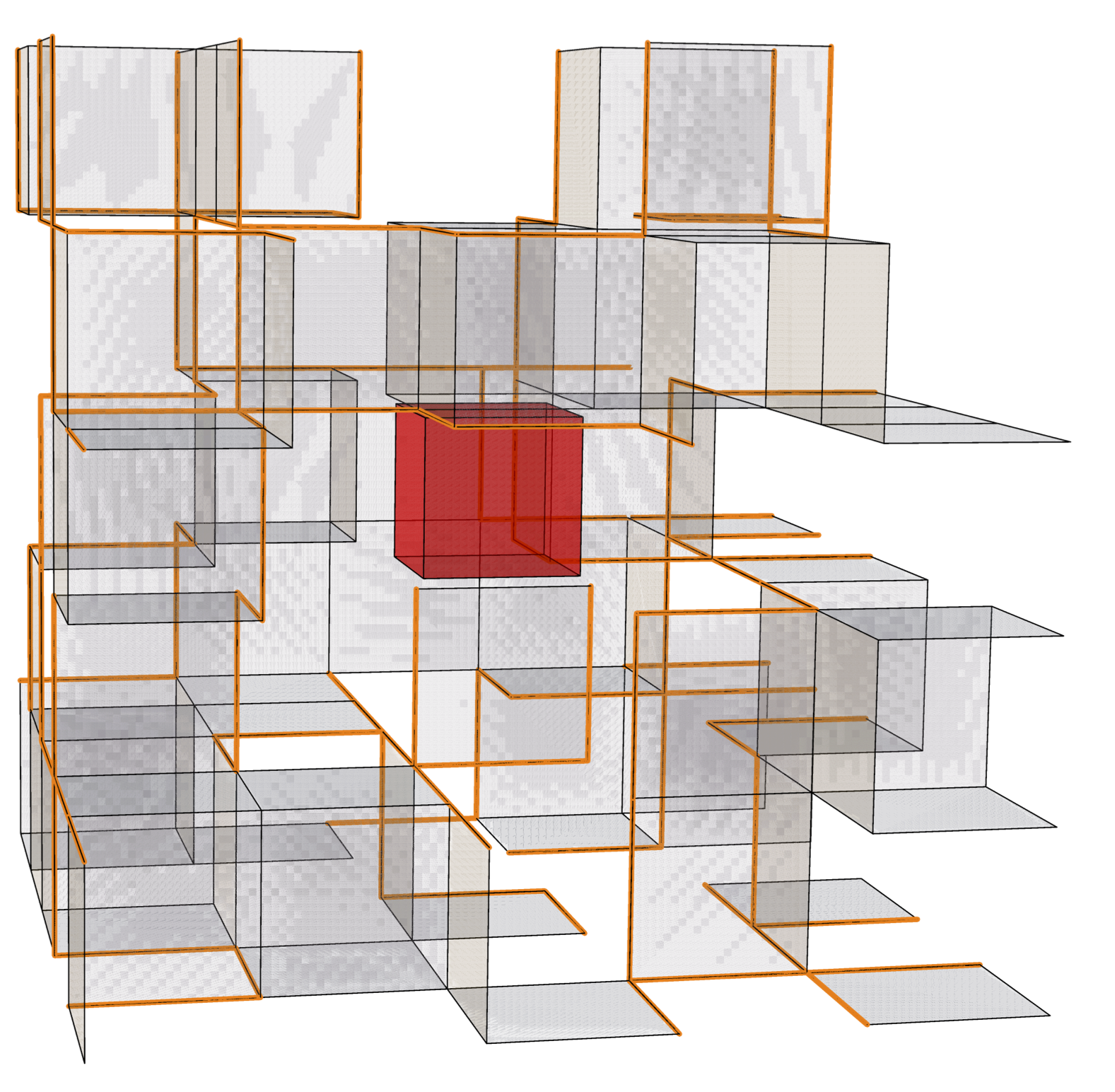}
    \caption{Simulation result of the string-wall network, based on the simple method in Ref.~\cite{Vachaspati:1984dz, Chang:1998tb}.
    We show a part where a closed wall emerges.
    A closed wall is highlighted in red. Gray translucent slices and orange lines represent open domain walls and strings, respectively.
    }
    \label{fig:string-wall-simu}
\end{figure}

One easy way to obtain the network structure is as follows~\cite{Vachaspati:1984dz, Chang:1998tb}.
Basically, the vacuum circle $\theta \in [0,2\pi]$ is divided into three equal-length parts: part I, centering at $0$; part II, centering at $2\pi/3$; and part III, centering at $4\pi/3$. After PQ symmetry breaking, each correlated patch randomly chooses one of three parts with equal probability. A string forms at the center if the surrounding correlated patches are assigned part numbers in clockwise (or anti-clockwise) order. A wall forms at the boundary of two correlated patches if they are assigned with II and III respectively. Consequently, the network contains two types of structures: open walls bounded by strings (each string connecting to a single piece of wall, $N_{\rm DW}=1$) and closed walls. We apply a large $N^3$ cubic lattice simulation in the realistic 3D case, and each cube represents a correlated Hubble patch. 
The simulation is purely a mathematical problem related to percolation theory, independent of specific axion parameters.
For illustrative purposes, we show the simulation results (part) in Fig.~\ref{fig:string-wall-simu}. Although simple, this simulation captures the key features: the network is dominated by open walls bounded by strings and a small portion is closed walls (``bubbles''). 

Then, we generalize the above simple model to a more realistic case where continuous variables are used in generating initial conditions; see \eg Refs.~\cite{Li:2023gil, Figueroa:2020rrl, Buschmann:2019icd, Hiramatsu:2010yn, Hiramatsu:2010yz}. 
The initial values of the PQ field $\Phi=(\phi_1+i \phi_2)/\sqrt{2}$, can be assigned based on the correlation functions in momentum space during inflation. By applying the Fourier transformation to $\phi_i$ in momentum space, one can obtain their values on each lattice site in the position space. The lattice spacing is taken as the correlation length, $\xi= 1/H_I$. 
$\theta =  {\rm arg}(\Phi)$ 
is continuous and domain walls occur at $\theta = \pi$. We focus on two parameters that can be derived from the statics of simulation: the area parameter $\mathcal{A}$ which is the number of walls per lattice cell on average, and $\gamma$, the ratio of total closed wall area to the total wall area. As we increase the simulation size $N^3$, $\mathcal{A}$ and $\gamma_a$ become stable. We adopt $N^3=512^3$, and we obtain
\begin{equation}
\begin{aligned}
\mathcal{A}\approx 37\%\pm0.007\%
,~~~
\gamma = 0.285\%\pm 0.002\%.\\
\end{aligned}
\end{equation}
with relatively small uncertainties. 
Also, most of the bubbles have a volume of one cell. Two-cell or larger bubbles are exponentially rare~\cite{1979PhR....54....1S} (even rarer with the strings present), which we neglect for now.
We refer the readers to the supplemental material (SUPP) for more simulation details.
\\

\noindent \textit{\textbf{PBH formation}}--We are going to demonstrate the criterion for a closed wall to collapse into a PBH.
The string-wall network, including walls bounded by strings and bubble walls, re-enters horizon at $T_{\rm en}$. The diameter of a bubble wall is $2R_0\simeq H^{-1}(T_{\rm en})$ identical to the distance between strings. A simple argument for PBH formation is that the bubble radius is smaller than its Schwarzchild radius $R_{S} = 2GM$~\cite{Vachaspati:2017hjw, Ge:2019ihf, Widrow:1989vj}. Bubbles lose energy by radiating free axions during collapse, which only becomes significant when their size $R\lesssim m_a^{-1}$~\cite{Fort:1993zb, Vachaspati:2017hjw, Ge:2019ihf}. Therefore, the only criterion we need for PBH formation is that, when the bubble shrinks to $R=m_a^{-1}$, it satisfies
\begin{equation}\label{eq:cri_estimate}
\begin{aligned} 
2G M \gtrsim m_a^{-1} ~~ \Rightarrow ~~
T_{\rm en}\lesssim 89.4~\MeV.
\end{aligned}
\end{equation}
We have used $M = 4\pi R_0^2 \sigma_w$ with the wall tension $\sigma_w = 9.23 f_a^2 m_a~$\cite{Sikivie:1982qv,Huang:1985tt}.  Also, we have used the fixed relation between axion mass $m_a$ and PQ energy scale $f_a$, $ \chi_0 = m_a^2 f_a^2$, where $ \chi_0\approx (75.6 ~\MeV)^4$ is the (zero-temperature) QCD topological susceptibility~\cite{Borsanyi:2016ksw}. The derived condition \eqref{eq:cri_estimate} is quite simple and independent of axion parameters. It only demands the network (therefore, the bubbles) re-enter horizon later than $89.4$~MeV. Note that bubble collapse indeed happens in this picture rather than expands because the gravitational repulsion of walls is negligible compared to the energy of enclosed cosmic plasma~\cite{Deng:2016vzb}.

Eq.~\eqref{eq:cri_estimate} is a rather conservative estimate. A much more accurate approach is to numerically solve the equation of motion of the closed axion wall field $a$~\cite{Vachaspati:2017hjw, Ge:2019ihf}. We define $E(r,t)$ as the energy enclosed within a radius $r$ at a time $t$ during collapse. Then, a PBH can form if $2GE(r, t) \gtrsim r$ is satisfied for some $r$ and $t$ during the collapse. Furthermore, we define $S(r,t)\equiv 2E(r,t)/r$, and $S_{\rm max}\equiv {\rm max}_{(r,t)} (2E(r,t)/r)$ is the maximum $S$ achieved during collapse. Therefore, PBH formation criterion can be equivalently expressed as $S_{\rm max}\gtrsim m_P^2$ where $m_P$ is the Planck mass. Our numerical results show that $S_{\rm max}$ is a linear function of the initial bubble radius $R_0$ in the logarithmic scale,
\begin{equation}\label{eq:cri}
\frac{S_{\rm max}}{f_a^2}
=
\begin{cases}
19.66 (m_{a}R_0)^{2.74}, &  T_{\rm en} \lesssim T_c \\
3.1\times10^3 [m_{a}(T_\text{en})R_0]^{2.76}, & T_{\rm en} \gtrsim T_c
\end{cases}.
\end{equation}
This is in the piecewise form because QCD axion mass is a function of $T$ which can be approximated as~\cite{Borsanyi:2016ksw},
\begin{equation}\label{eq:axion_mass_T}
m_a(T) =
\begin{cases}
\chi_0^{1/2}/f_a, &  T \lesssim T_c \\
\chi_0^{1/2}/f_a \cdot (T/T_c)^{-n}, & T  \gtrsim T_c
\end{cases}.
\end{equation}
One of the most updated calculations~\cite{Borsanyi:2016ksw} indicates that $T_c\approx 150 \MeV$ near QCD transition and $n\simeq 4$. Initially, axion mass increases rapidly as temperature drops. After $T_c$, it settles to constant $m_a$.

\begin{figure}[tb]
    \centering
    \includegraphics[width=1\linewidth]{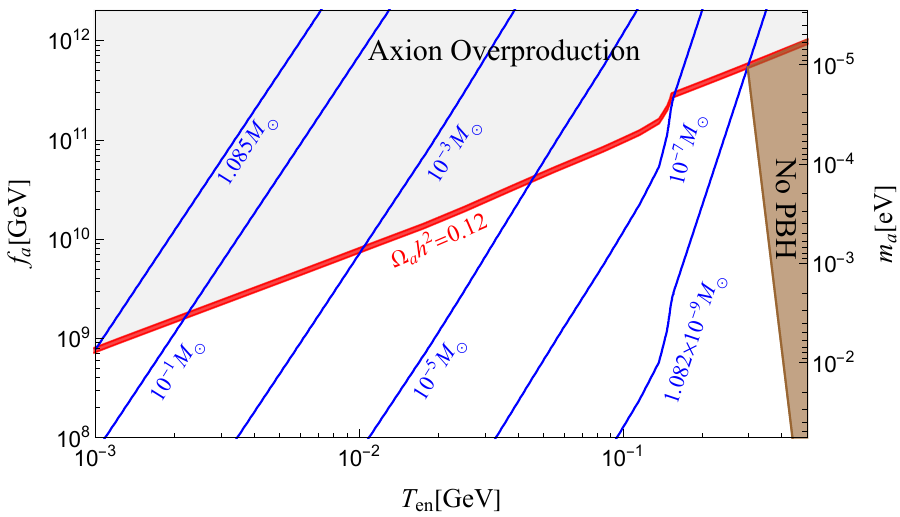}
    \caption{Parameter space for PBH formation. The shaded brown region is the parameter space where PBH cannot form. Additionally, the red band represents that free axions from the decay of walls bounded by strings fully account for dark matter. The shaded gray region is excluded because of dark matter overproduction. The blue lines are the PBH mass contours.
    }
    \label{fig:PBH-Tent}
\end{figure}

Our result~\eqref{eq:cri} is consistent with that in Refs.~\cite{Vachaspati:2017hjw, Ge:2019ihf}, except a small discrepancy between the prefactor in the first branch and that in Ref.~\cite{Vachaspati:2017hjw} because we have included the Universe's expansion in solving the bubble dynamics. Plugging Eq.~\eqref{eq:cri} into the criterion $S_{\rm max}\gtrsim m_P^2$, we get the parameter space of $(T_{\rm en}, f_a)$ that allows PBH formation, shown in Fig.~\ref{fig:PBH-Tent}. Also, the resultant PBH mass can be estimated as $M_{\rm PBH}\simeq 4\pi R_0^2 \cdot \sigma_w(T_{\rm en}) $. From Fig.~\ref{fig:PBH-Tent}, PBH mass range in this scenario is 
\begin{equation}\label{eq:mass_range}
10^{-9}~M_{\odot} \lesssim M_{\rm PBH} \lesssim 1 ~ M_{\odot} .
\end{equation}
The later the network re-enters horizon, the heavier the resultant PBH will be. 
The lower bound in Eq.~\eqref{eq:mass_range} corresponds to
$T_{\rm en} \sim 300~\MeV$, below which PBH can form, 
while the higher bound corresponds to the lowest temperature of re-entering horizon, $T_{\rm en} \sim 1~\MeV$, which is required by not disturbing BBN observations
\footnote{This bound may be slightly modified by studying carefully how non-zero $\theta$ affects BBN processes. See also the BBN constraints in a different context in \eg Refs.~\cite{Kawasaki:1999na, Kawasaki:2000en, Hasegawa:2019jsa}.}. 
This mass range~\eqref{eq:mass_range} is drawn for the case that free axions from the decay of open walls bounded by strings explain the dark matter relic abundance (see Eq.~\eqref{eq:axion_abundance} below).
Note that PBHs obtained here are light, so their mass accretion after formation is negligible (see \eg Ref.~\cite{Deng:2016vzb}).
If free axions only account for a fraction of dark matter, lighter PBHs can form, and the PBH formation criterion can be extended up to 
$T_{\rm en} \sim 400~\MeV$
where it reaches the constraint $f_a \gtrsim 10^{8}~\GeV$ from Supernova 1987A~\cite{Chang:2018rso,Carenza:2019pxu}, as can be seen in Fig.~\ref{fig:PBH-Tent}.  
\\

\noindent \textit{\textbf{PBH abundance}}--
In addition to closed walls, walls bounded by strings collapse immediately after $T_{\rm en}$, since strings are pulled toward each other by the wall tension. This is because wall energy dominates the system, $\sigma H^{-1}(T_{\rm en})/f_a^2 \sim m_a H^{-1}(T_{\rm en}) \gg 1$. Such collapse of walls bounded by strings releases energy mainly in the form of free axions\footnote{
A tiny portion of energy is released as gravitational waves (GWs). This generates GW spectra drastically different from that in the network scaling regime~\cite{Ge:2023rce}. For certain parameter space of axion-like particles, such GW spectra can possibly explain the reported nano-Hertz stochastic GW background and can be tested by various GW interferometry experiments~\cite{Ge:2023rce}.
},
with the energy density estimated as the wall energy density, $\rho_{a}(T_{\rm en}) \simeq \rho_{w}(T_{\rm en}) \simeq \mathcal{A} \sigma(T_{\rm en}) H(T_{\rm en})$. Furthermore, numerical simulations show that the resultant free axions are mildly relativistic, with a Lorentz factor $\gamma_a \approx 3.23\pm0.18$~\cite{Hiramatsu:2012gg, 
Kawasaki:2014sqa}. 
Thus, the present-day energy density of free axions is 
\begin{equation}
\begin{aligned}
    \rho_a(T_0)
    &= \frac{1}{\gamma_a}\rho_a(T_{\rm en}) \frac{a^3( T_{\rm en})}{a^3(T_0)} \\
    &\simeq
    9.23
    \frac{\mathcal{A}}{\gamma_a}
     \chi_0^{1/2}f_a
    H(T_{\rm en})
    \frac{a^3( T_{\rm en})}{a^3(T_0)}.
    \end{aligned}
\end{equation}
$a$ is the scale factor. In deriving the last step, we have used the relations $\sigma_w = 9.23 f_a^2 m_a~$\cite{Sikivie:1982qv,Huang:1985tt} and $ \chi_0 = m_a^2 f_a^2$ for QCD axion.
Then, the present-day ratio of the axion energy density to the critical $\rho_{\rm cri}$ can be calculated as
\begin{equation} \label{eq:axion_abundance}
\begin{aligned}
\Omega_{a} h^2 &=
\frac{\rho_a(T_0)}{\rho_{\rm cri}(T_0)} h^2
\\
&\simeq 
0.068\frac{\mathcal{A}}{\gamma_a} \left[\frac{10.75}{g_{*}(T_{\rm en})}\right]^{\frac{1}{2}}
\left( \frac{f_a}{10^9~\GeV} \right) 
\left( \frac{20~\MeV}{T_{\rm en}} \right).
\end{aligned}
\end{equation}
$g_{*}(T_{\rm en})$ is the effective numbers of relativistic degrees of freedom at $T_{\rm en}$.
Compared with Ref.~\cite{Harigaya:2022pjd}, we have included $\gamma_a$ and also the area parameter $\mathcal{A}$. Axions can fully account for the dark matter if $\Omega_a h^2 =0.12$, corresponding to the red band in Fig.~\ref{fig:PBH-Tent}. In comparison, the misalignment mechanism contributes in this scenario with the number density $n_{a,{\rm mis}}(T_1) \sim m_a(T_1) f_a^2  \left<\theta^2\right>$, which is negligible since $m_a(T_{\rm en})n_{a,{\rm mis}}(T_{\rm en})/\rho_{a}(T_{\rm en}) \sim 0.1 T_{\rm en}/T_1 \ll 1$.

\begin{figure}
    \centering
    \includegraphics[width=1\linewidth]{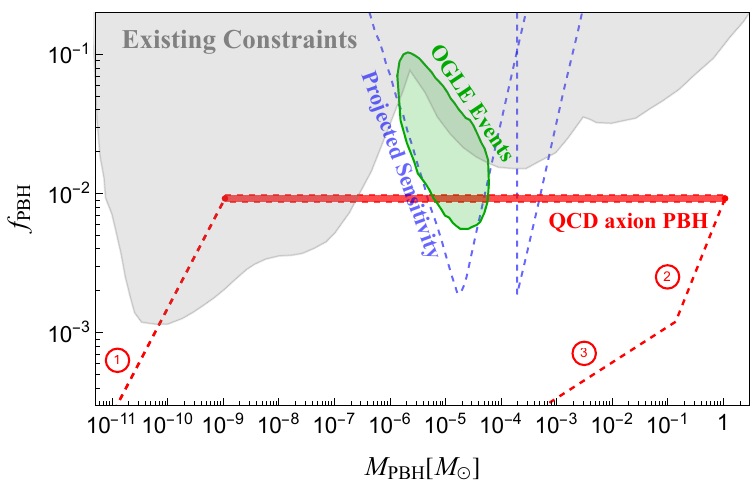}
    \caption{Plot of PBH abundance and its monochromatic mass. The solid red band denotes $f_{\rm PBH} \simeq 0.92\%\pm0.05\%$, when the dark matter relic abundance is accounted for by free axions from open walls bounded by strings. The corresponding mass range is given in Eq.~\eqref{eq:mass_range}. The three labeled dashed red lines are mapped from the requirements shown in Fig.~\ref{fig:PBH-Tent}, which are: 
    {\normalsize \ding{172}} PBH formation criterion; {\normalsize \ding{173}} $T_{\rm en}\gtrsim 1~\MeV$; and {\normalsize \ding{174}} $f_a\gtrsim 10^8~\GeV$. 
    The shaded elliptic region represents the PBH explanation to the OGLE events~\cite{Niikura:2019kqi, 2017Natur.548..183M}.
    The existing PBH constraints in the shaded gray region mainly come from the lensing~\cite{Niikura:2019kqi, 2017Natur.548..183M, EROS-2:2006ryy, Niikura:2017zjd} under the null hypothesis that there is no PBH. The projected sensitivities are from the future ET by observing PBH binaries~\cite{Miller:2020kmv}. 
    }
    \label{fig:f-PBH-search}
\end{figure}

We can easily express the PBH abundance as
\begin{equation}
f_{\rm PBH} \equiv \frac{\Omega_{\rm PBH}}{\Omega_{\rm DM}} = \gamma \cdot \gamma_a \frac{\Omega_a}{\Omega_{\rm DM}}.
\end{equation}
If axions from open walls can fully explain dark matter, $\Omega_a = \Omega_{\rm DM}$, then PBH abundance is fixed as $f_{\rm PBH} =\gamma \gamma_a = 0.92\%\pm 0.05\%$. We emphasize that $\gamma$ is insensitive to axion parameters or $T_{\rm en}$. It is determined by the ratio of closed walls to total walls, which is a constant predicted by percolation theory.
The possible uncertainty from $\gamma_a$ simulations~\cite{Hiramatsu:2012gg, Kawasaki:2014sqa} has been included. Our PBH mass spectrum is monochromatic, since closed walls have a uniform volume, enclosing a single cube of correlated patch.
We show $f_{\rm PBH}$ in Fig.~\ref{fig:f-PBH-search}, together with the constraints on $f_a$, $T_{\rm en}$, and also the PBH formation criterion mapped from Fig.~\ref{fig:PBH-Tent}.

In comparison, we show the existing constraints on $f_{\rm PBH}$ based on gravitational microlensing observations~\cite{Niikura:2019kqi, 2017Natur.548..183M, EROS-2:2006ryy, Niikura:2017zjd}. Besides, GWs generated from PBH binaries can be probed in principle by GW detectors, such as LIGO-Virgo \cite{Chen:2021nxo}, Tianqin \cite{Liu:2020eko}, Taiji \cite{Yang:2022cgm} and LISA \cite{Yang:2022cgm}, etc. For our planetary-mass PBHs, they can be potentially probed by the prospective Einstein Telescope (ET) via searching for GWs from binaries in our galaxy and in the solar system vicinity~\cite{Miller:2020kmv}. 

More interestingly, OGLE collaboration has observed 6 ultrashort-timescale microlensing events~\cite{2017Natur.548..183M} which can be well reproduced by PBHs~\cite{Niikura:2019kqi}. 
As we can see from Fig.~\ref{fig:f-PBH-search}, PBHs required to explain such events partly overlap with the PBHs generated in our mechanism.
Hence, from a positive perspective, observations by OGLE~\cite{Niikura:2019kqi, 2017Natur.548..183M} might have potentially revealed supporting evidence for the existence of PBHs proposed in this work.

Another intriguing possibility is that Planet 9 in the solar system can be a PBH with mass $\sim 10^{-5} M_{\odot}$, which explains the growing observational anomalies of trans-Neptunian objects' orbits~\cite{Scholtz:2019csj}. Such a PBH can be naturally generated in our mechanism.  
This idea is testable via searching for signals from PBH interactions with the surrounding environment~\cite{Scholtz:2019csj, Siraj:2020upy} or a wilder proposal of sending a small laser-launched spacecraft to approach the PBH~\cite{Witten:2020ifl}.

Lastly, it should be noted that the numerical values $\gamma_a$ are different given by different groups~\cite{Hiramatsu:2012gg, Kawasaki:2014sqa, Chang:1998tb, Harigaya:2022pjd}.
The string core's size is $f_a^{-1}$ while the wall thickness is $\sim m_a^{-1}$. This multi-scale ($m_a/f_a\sim 10^{-25}$) makes it difficult for lattice simulations of the dynamical collapse of open walls bounded by strings~\cite{Chang:1998tb}.
One of the most updated simulations regarding $\gamma_a$~\cite{Hiramatsu:2012gg, Kawasaki:2014sqa} gives $\gamma_a \approx 3.23\pm 0.18$ by extrapolating the simulation results at the artificial value of $\Lambda_{\rm QCD}/f_a = 0.275\sim 0.4$ to the physical value of $10^{-12}$ ($\Lambda_{\rm QCD}\sim 400$~MeV). We have adopted this $\gamma_a$ in making Figs.~\ref{fig:PBH-Tent} and~\ref{fig:f-PBH-search}, but one can simply shift the results if using other values of $\gamma_a$. 
For example, an old simulation~\cite{Chang:1998tb} obtains $\gamma_a\simeq 60$ by extrapolating the simulation results at the artificial $\ln(f_a/m_a)\sim 4.6$ to the physical $\ln(f_a/m_a)\sim 60$. This would give 
$f_{\rm PBH}=\gamma\gamma_a \sim 17\%$
and the PBH mass range as
$\sim 4\times10^{-8}-21 M_{\odot}$.
One caveat we would like to point out is that Refs.~\cite{Hiramatsu:2012gg, Kawasaki:2014sqa, Chang:1998tb} carried out simulations for the traditional post-inflationary scenario, while our during-inflation scenario is different where the network re-enters horizon and collapse at a much later time, $m_a\gg H(T_{\rm en})$.
Therefore, it is possible that the level of uncertainty in calculating the PBH abundance caused by $\gamma_a$ has been underestimated. 
Besides, Ref.~\cite{Harigaya:2022pjd} simply assumes the typical momentum of free axions to be the Hubble scale, $H(T_{\rm en})$, and therefore $\gamma_{a}\simeq 1$. This would give 
$f_{\rm PBH}=\gamma\gamma_a \sim 0.3\%$ 
and the PBH mass range as
$\sim 2\times10^{-10}-0.35M_{\odot}$.
Simulation of $\gamma_a$ is well beyond the topic of PBH formation in this work, since it is related to open walls bounded by strings while PBHs are from closed walls.
\\

\noindent \textit{\textbf{Conclusions}}--
A new possible mechanism is introduced for producing PBHs within the QCD axion framework. In this mechanism, PQ symmetry breaks during inflation, so the $N_{\rm DW} = 1$ axion topological defects re-enter horizon and collapse at a temperature sufficiently later than the post-inflationary case. Closed axion domain walls are sufficiently large to collapse into PBHs. Numerical lattice simulation shows that 
$\sim0.3\%$
of walls are closed walls. This outcome remains independent of axion parameters and the re-entering horizon temperature $T_{\rm en}$, with its determination rooted in the mathematical principles of percolation theory.

Dark matter is accounted for by free axions from the collapse of open walls bounded by strings. If we adopt the Lorentz factor of free axions simulated by Refs~\cite{Hiramatsu:2012gg, Kawasaki:2014sqa},
our PBHs constitute 
$\sim 0.9\%$ 
of dark matter, independent of $T_{\rm en}$.
The resultant PBH mass is monochromatic, which is between $\sim10^{-9}-1 M_{\odot}$ for the classical QCD axion window
$m_a\sim 10^{-5}-10^{-2}$~eV, 
depending on 
$T_{\rm en} \sim 300-1~\MeV$.
If axions only partially account for dark matter, more parameter space is released, allowing lighter PBHs to form and extending the PBH formation criterion to 
$T_{\rm en} \lesssim 400$~MeV.

We have also discussed some cosmological and astronomical implications. 
Without fine-tunings, PBHs produced in this mechanism can naturally explain the microlensing events observed by OGLE~\cite{Niikura:2019kqi, 2017Natur.548..183M}. They could also be a candidate for the PBH-origin Planet 9 in our solar system~\cite{Scholtz:2019csj, Witten:2020ifl}. Additionally, binaries of such PBHs can be potentially probed by GW detectors such as the prospective Einstein Telescope~\cite{Miller:2020kmv}.\\

\begin{acknowledgments}
\noindent \textit{\textbf{Acknowledgment}}--We would like to thank Haipeng An, Heling Deng, Bin Guo, Huai-Ke Guo, Yong Tang, and Lian-Tao Wang for useful discussions. The work of SG is supported by the National Natural Science Foundation of China (NSFC) under Grant No. 12247147, the International Postdoctoral Exchange Fellowship Program, and the Boya Postdoctoral Fellowship of Peking University.
The work of JL is supported by NSFC under Grant No. 12075005, 12235001. 
  
\end{acknowledgments}

~\\
~\\
~\\
\appendix

\section{Supplemental Material}
This Supplemental Material provides more details of the model and simulations.


\subsection{A. Equations of Motion}\label{sec:app-A}

As described in the main text, the Lagrangian of the model can be written as
\begin{equation}\label{eq:supp_lagrangian}
    \begin{aligned}
        \mathcal{L} = \frac{1}{2}(\partial\Phi)^2 - V(\Phi)
        ,~
        V(\Phi) = \frac{\lambda}{4}(\left| \Phi \right|^2 - v_a^2)^2 + c \phi^2 \left| \Phi \right|^2.
    \end{aligned}
\end{equation}
$c$ is the coupling between the PQ field $\Phi$ and the inflaton $\phi$. The inflation drives the PQ symmetry breaking when $\phi$ rolls down to the value of $\sqrt{\lambda/2c}v_a$.

After the PQ symmetry breaks during inflation, the initial conditions will be stretched out of the horizon by inflation and become homogenized over enormous distances. The system is kept frozen until the network re-enters the horizon. This is valid as long as the spatial derivative is ignorable. This can be seen from the equation of motion of the PQ field $\Phi=(\phi_1 + i\phi_2)/\sqrt{2}$:
\begin{equation}
\begin{aligned}
    \ddot{\phi}_i+3H\dot{\phi}_i-\frac{1}{a(t)^2} \nabla^2 \phi_i + \frac{\partial V}{\partial\phi_i}=0,
\end{aligned}
\end{equation}
$a(t)$ is the scale factor and $\nabla$ is the derivative with respect to the co-moving spatial coordinates. 
At the PQ symmetry breaking, the correlation length is $\sim H_I$, so we can take the initial values in a Hubble radius to be uniform, and they will be further stretched out of horizon exponentially due to inflation. Due to this homogeneity, the spatial derivative suppressed by $a(t)^{-2}$ can be neglected during the evolution out of horizon. Later on, the axion mass $m_a$ effectively turns induced by non-perturbative QCD effects which becomes $m_{a}\sim H$ at $\sim 1$~GeV. This overcomes the Hubble friction proportional to $3H$, and the axion field in one homogeneous patch (still out of horizon) starts to roll down (misalignment mechanism) either to 0 or $2\pi$ determined by the initial values in our simulations. 
Our final results may be modified by the following effects: the closed walls are not perfectly spherical, the closed walls re-enters the horizon not exactly at the same time, etc. However, these effects are subdominant and will not destroy the robust steps of forming PBHs described in the main text.



Next, we comment on the generality of the model.
For the PQ symmetry breaking to be driven by inflation, the assumption here is that $\Phi$ couples to the $\phi$. One simple realization is via $c\phi^2 \Phi^\dagger\Phi$ as shown in Eq.~\eqref{eq:supp_lagrangian}. Such coupling has been widely discussed in the literature. 
It dates back to early literature of axion cosmology; see e.g., Ref.~\cite{Linde:1991km} (Section 4.3), which indicated that this coupling is \textit{more realistic} to occur than the case of no coupling.
Ref.~\cite{Linde:1991km} also discussed the case that the PQ symmetry breaks during inflation at $\phi = \phi_c = \sqrt{\lambda/2c} v_a$, which is exactly what we need for our setting
\footnote{Note the terminology difference. In their terminology, it is $\phi = m_{\Phi}/\sqrt{\nu}$ ($\nu$ is the coupling) and $m_{\Phi} = \sqrt{\lambda/2} v_a$.}. 
It implies that $v_a$ is significantly larger than $H_I$ to suppress the thermal correction, which ensures that the symmetry breaking is driven by the inflaton field at $\phi_c$.
In addition, we naturally benefit from the fact that the subsequent axion fluctuation $\propto H/v_a$ is suppressed.  
Therefore, we make no non-trivial assumptions beyond the classical setting in Ref.~\cite{Linde:1991km}. 
In addition to Ref.~\cite{Linde:1991km}, the couplings between the PQ field and the inflaton field, represented by $c\phi^2 \Phi^\dagger\Phi$ and its variants, have also been commonly explored in many different contexts and for many different purposes; see \eg Refs.~\cite{Rosa:2021gbe, Redi:2022llj, Bao:2022hsg, Harigaya:2015hha, Kearney:2016vqw, Jeong:2013xta, An:2022cce, An:2023idh, Ismail:2024zbq}. 
Our novelty lies in the PBH formation within this basic framework.

Additionally, our model can be realized not necessarily by the coupling above. For example, the inflationary Hubble scale $H_I$ may also drive the PQ symmetry breaking if it varies significantly during inflation (see e.g., Ref.~\cite{Redi:2022llj}). This may be realized via the coupling between $\Phi$ and the Ricci scalar, $-\xi R \Phi^{\dagger}\Phi = 2H_I^2 \Phi^{\dagger}\Phi$, where the Ricci scalar $R=-12 H_I^2$ and a natural choice of $\xi$ is $1/6$ with conformal symmetry imposed.

\subsection{B. Simulations}

We extend the simulation with three discrete variables~\cite{Vachaspati:1984dz, Chang:1998tb} to a more realistic one with continuous variables, by adopting a well-studied lattice simulation of strings and domain walls based on, for example, Refs.~\cite{Li:2023gil, Figueroa:2020rrl, Buschmann:2019icd}. In our scenario, strings and initial values are stretched and blown out of the horizon during the inflation. As a result, the spatial dynamics become less significant, as shown in the section above. 
As a result, the general two-stage simulations (PQ era and QCD era) become one-stage, where only the initial values of the field configuration in the PQ era play an important role.

We write the PQ field as $\Phi = (\phi_1+i\phi_2)/\sqrt{2}$. During inflation, the amplitude distribution of the real scalars, $\phi_1$ and $\phi_2$, in momentum space at the finite temperature $T_I=\frac{H_I}{2\pi}$ can be derived as,
\begin{equation}\label{eq:therm-dis}
\begin{aligned}
    \mathcal{P}_{\phi_i}(k)
    &=\frac{1}{2a(t)^3\omega_k}\left[1+\left(\frac{H_I}{\omega_k}\right)^2\right]\left(1+2n_k \right).
\end{aligned}
\end{equation}
$\omega_k=\sqrt{k^2/a(t)^2+m_{\rm eff}^2}$ where $k$ is the four-momentum and $a(t)$ is the scale factor. $m_{\rm eff}=\sqrt{2c\phi^2 -\lambda v_a^2}$ is the effective mass of the scalars $\phi_1$ and $\phi_2$. When the inflaton field $\phi$ rolls down to the value of $\sqrt{\lambda/2c}v_a$, PQ symmetry breaks. We have omitted the finite-temperature correction of $m_{\rm eff}^2$ which is $\sqrt{\lambda T_I^2}/3$. This is because $T_I\ll v_a$ such that the inflation drives the symmetry breaking while the temperature is less important~\cite{Harigaya:2022pjd}. $n_k$ is the Bose-Einstein distribution at $T_I$. The correlation function of the PQ fields $\phi_i$ in the momentum space can be written as
\begin{equation}
    \langle \phi_i(\textbf{k})\phi_j(\textbf{k}') \rangle = (2\pi)^3\mathcal{P}_{\phi_i}(k)\delta(\textbf{k-k}')\delta_{ij}.
\end{equation}
In the discrete space, this can be expressed as~\cite{Li:2023gil, Figueroa:2020rrl}
\begin{equation}
    \langle| \phi_i(\textbf{k}))|^2 \rangle = \left( \frac{N}{\Delta x_{\rm phy}} \right)^3\mathcal{P}_{\phi_i}(k), ~~\langle \phi_i(\textbf{k}) \rangle =0.
\end{equation}
where $N^3$ is the lattice size and $\Delta x_{\rm phy}$ is the physical lattice spacing.
We then can generate the Gaussian random distribution of $\phi_i(\textbf{k})$ in momentum space with the variance as $\langle| \phi_i(\textbf{k}))|^2 \rangle$. Physically, the momentum $\textbf{k}$ suffers from the IR and UV truncation based on the lattice, which requires that $|\textbf{k}| \in [ \frac{2\pi}{N \cdot \Delta x_{\rm phy}}, \frac{\sqrt{3}\pi}{\Delta x_{\rm phy}}]$ \cite{Figueroa:2020rrl}. 
Besides, one also needs to generate two random phases that follow uniform distribution in the range $[0,2\pi)$  (For more details, one can refer to Sec. 7.1 in Ref.~\cite{Figueroa:2020rrl}).
Finally, by applying the discrete Fourier transform to the field in momentum space, one can obtain the initial values of the scalar fields $\phi_i$ in the real space.

With the initial values of $\phi_1$ and $\phi_2$ at each lattice site known, one can calculate the angular variable $\theta$ (axion field rescaled by $v_a$) at the site as 
\begin{equation}
\theta =\arg(\Phi)=
\begin{cases}
    \arctan\frac{\phi_2}{\phi_1},~~~~~~~~\phi_1>0 ~{\rm and}~ \phi_2>0; \\
    \arctan\frac{\phi_2}{\phi_1}+\pi,~~~\phi_1<0;\\
    \arctan\frac{\phi_2}{\phi_1}+2\pi,~~\phi_1>0 ~{\rm and}~ \phi_2<0.\\
\end{cases}
\end{equation}
Domain walls will form at the positions of $\theta = \pi$. Equivalently speaking, a domain wall intersects the link between two sites if $\phi_1<0$ and $\phi_2$ has different signs at the two sites (see e.g., Refs.~\cite{Li:2023gil, Hiramatsu:2010yn}). A closed wall can form if a site and all of its neighboring sites satisfy this condition.
Following this method, one can count the number of walls and closed walls, from which one can further obtain the area parameter $\mathcal{A}$ and the ratio of the total closed wall area to the total wall area, $\gamma$. 

There are some differences between our simulations and those in the post-inflationary scenario (PQ symmetry breaks after inflation). In the post-inflationary scenario, the simulations usually set the power spectrum of $\phi_i$ as~\cite{Hiramatsu:2010yz}:
\begin{equation}\label{eq:fluctuation_post_inflation}
\begin{aligned}
    \mathcal{P}_{\phi_i}(k) =\frac{1}{2a(t)^3\omega_k}\left[1+2n_k\right].
\end{aligned}
\end{equation}
In comparison, in our during-inflation scenario, the power spectrum Eq.~\eqref{eq:therm-dis} has an extra term $(H_I/\omega_k)^2 = \left(H_I/\sqrt{k^2/a(t)^2+m^2}\right)^2$ from the superhorizon effect which becomes important during inflation. 
In fact, this effect is very common in dealing with the initial fluctuations during inflation (see e.g., Chapter 8 in Ref.~\cite{Mukhanov:2005sc}). For subhorizon, $k>H_I$, Eq.~\eqref{eq:therm-dis} becomes 
\begin{equation}
\begin{aligned}
    \mathcal{P}_{\phi_i}(k)\simeq\frac{1}{2a(t)^3\omega_k}\left[1+2n_k\right].
\end{aligned}
\end{equation}
which is the same as Eq.~\eqref{eq:fluctuation_post_inflation}. While for superhorizon, $k<H_I$, Eq.~\eqref{eq:therm-dis} becomes
\begin{equation}
\begin{aligned}
    \mathcal{P}_{\phi_i}(k)\simeq\frac{H_I^2}{2a(t)^3\omega_k^3}\left[1+2n_k\right].
\end{aligned}
\end{equation}
The physical meaning is clear, especially in the position space. For the case within the horizon, the correlation function, $\langle \Phi(x)\Phi(y) \rangle$, decreases with the increase of their distance, as if in the Minkowski space.
When the distance exceeds the horizon length, the correlation function is determined by this Hubble length scale. For the traditional post-inflationary scenario, the typical momentum of the fluctuation is $k\sim T$, which is much larger than the Hubble constant $H\propto T^2/M_P$. As a result, the superhorizon term can be ignored. While for our case where $T_{I} \sim H_I$, the distance between two points will be persistently stretched out of horizon much more efficiently by the e-folding expansion of inflation, so the superhorizon term cannot be neglected. As the inflaton field slowly rolls down, PQ symmetry breaks as $m_{\rm eff}=\sqrt{2c\phi^2 -\lambda v_a^2}$ approaches zero. Therefore, at the phase transition, $m_{\rm eff}\lesssim H_I$ and the correlation is determined by $H_I^{-1}$ rather than $m_{\rm eff}^{-1}$, and the lattice spacing can be properly set as $H_I^{-1}$.

\begin{figure}[htbp]
    \centering
    \includegraphics[width=0.8\linewidth]{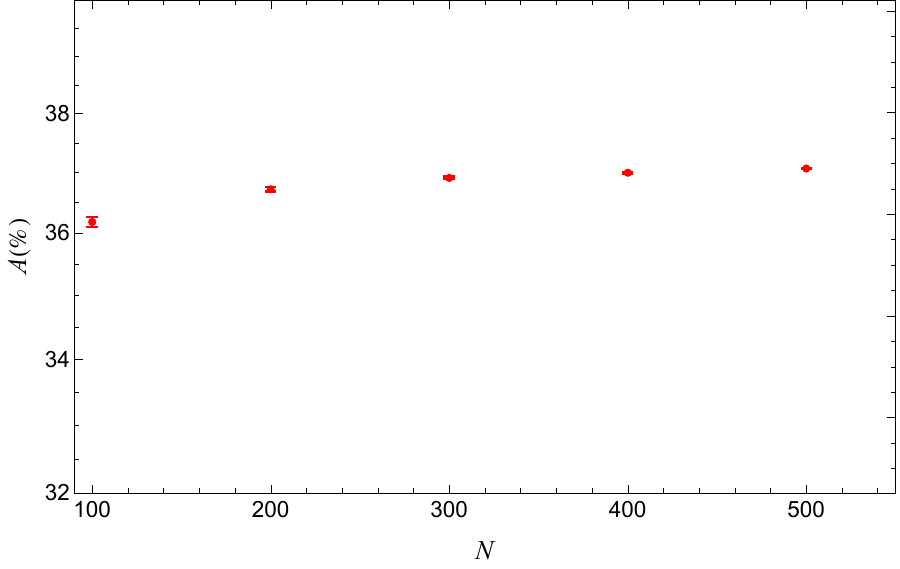}\\
    \includegraphics[width=0.8\linewidth]{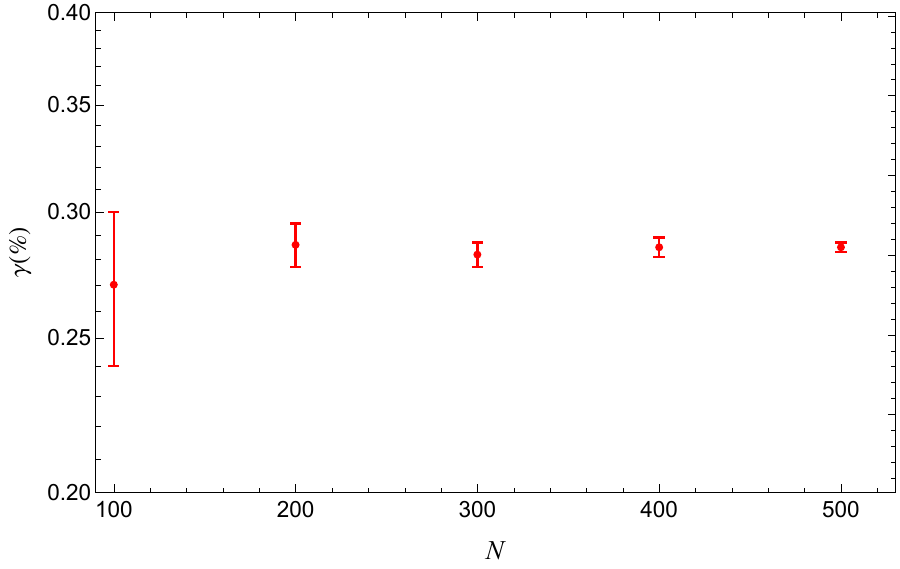}
    \caption{
    Simulation results of $A$ and $\gamma$. The lattice size is $512^3$.
    }
    \label{fig:simu-resu-B}
\end{figure}

We perform the simulations in different lattice sizes starting from $N^3=100^3$. For each size, we repeat the simulation ten times to get the uncertainties of $\mathcal{A}$ and $\gamma$. The results are shown in Tab.~\ref{tab:resl-B} and Fig.~\ref{fig:simu-resu-B} with error bars. We can see that both the values of $\mathcal{A}$ and $\gamma$ become stable as the lattice size increases, with the uncertainties dropping progressively. Finally, we adopt the values at $N^3=512^3$ which are $\mathcal{A}\approx 37\%$ and $\gamma\approx 0.285\%$ with relatively small uncertainties. 

\begin{table}[htbp]
    \centering
    \begin{tabular}{c|cc}
    \hline
       $N^3$  & $\mathcal{A}(\%)$ & $\gamma(\%)$ \\
       \hline
        $100^3$ & $36.18\pm0.08$ & $0.27\pm0.03$\\
        $200^3$ & $36.72\pm0.04$ & $0.286\pm0.009$\\
        $300^3$ & $36.91\pm0.02$ & $0.282\pm0.005$\\
        $400^3$ & $36.996\pm0.009$ & $0.285\pm0.004$\\
        $512^3$ & $37.066\pm0.007$ & $0.285\pm0.002$\\
    \hline
    \end{tabular}
    \caption{
    Simulation results of $\mathcal{A}$ and $\gamma$ for different lattice sizes.
    }
    \label{tab:resl-B}
\end{table}

Next, we comment on the size distribution of closed walls.
The percolation predicts that the number of clusters (closed walls) decreases exponentially with the size increasing~\cite{1979PhR....54....1S}. Large clusters are even rarer with the strings present. 
The probability for a lattice cell to be occupied by a closed wall is $\sim \gamma$. Then, the probability for a two-cell closed wall to occur can be simply estimated as $\sim \gamma^2$ which is much smaller.
Therefore, the number density of closed walls larger than one cell is tiny, which has been neglected in our further discussion. The PBH mass spectrum is monochromatic because the lattice cell is uniform in volume, which is an artificial effect. Indeed, in the realistic case, the spectrum is not perfectly monochromatic. We expect the volume of closed walls to follow a Gaussian-like distribution centering at one cell. The dispersion (variance) of the distribution should be small due to the following reasons. On the one hand, the initial sizes of closed walls bigger than one cell are rare, and the network dynamics during inflation are trivial which does not affect the initial conditions.
On the other hand, after re-entering horizon, the collapse of open walls bounded by strings would not significantly alter the dynamics of closed walls, as these two kinds of objects are separate and there are no strings intersecting with closed walls.
Also, the collapse is very quick, which can be completed in about one Hubble time, $H(T_{\rm en})^{-1}$, which has less effect on the initial bubble size distribution compared to the post-inflationary scenario with a scaling-regime evolution. In this sense, the PBH mass in
Fig.~\ref{fig:PBH-Tent} and Fig.~\ref{fig:f-PBH-search} 
in the main text
should be regarded as the average mass, although the dispersion is narrow which is close to a monochromatic distribution.

\bibliographystyle{utphys}
\bibliography{references}
\clearpage


\clearpage
\end{document}